\RequirePackage{fix-cm}
\documentclass[twocolumn]{svjour3}          

\smartqed  
\usepackage{graphicx}
\usepackage{amsmath}
\usepackage[ruled]{algorithm2e}
\usepackage{fixltx2e}
\usepackage[export]{adjustbox}
\usepackage{diagbox}
\usepackage{booktabs}
\usepackage{caption}

\begin{document}

\title{Optimal Virtualized Inter-Tenant  Resource Sharing for Device-to-Device Communications in 5G Networks
}


\author{Christoforos Vlachos     \and
        Vasilis Friderikos		 \and
        Mischa Dohler
}

\institute{C. Vlachos (corresponding author), \textit{Student Member IEEE} \at
              Centre for Telecommunications Research, Department of Informatics \\ King's College London, UK \\
              \email{christoforos.vlachos@kcl.ac.uk}           
           \and
           V. Friderikos, \textit{Member IEEE} \at
              Centre for Telecommunications Research, Department of Informatics \\
              King's College London, UK 
           \and
          M. Dohler, \textit{Fellow IEEE} \at
              Centre for Telecommunications Research, Department of Informatics \\
              King's College London, UK 
}

\date{Received: date / Accepted: date}

\maketitle

\begin{abstract}
Device-to-Device (D2D) communication is expected to enable a number of new services and applications in future mobile networks and has attracted significant research interest over the last few years. Remarkably, little attention has been placed on the issue of D2D communication for users belonging to different operators. In this paper, we focus on this aspect for D2D users that belong to different tenants (virtual network operators), assuming virtualized and programmable future 5G wireless networks. Under the assumption of a cross-tenant orchestrator, we show that significant gains can be achieved in terms of network performance by optimizing resource sharing from the different tenants, i.e., slices of the substrate physical network topology. To this end, a sum-rate optimization framework is proposed for optimal sharing of the virtualized resources. Via a wide site of numerical investigations, we prove the efficacy of the proposed solution and the achievable gains compared to legacy approaches.
\end{abstract}

\section{Introduction}
\label{intro}
The Device-to-Device (D2D) communication paradigm is expected to become the key enabler for propelling local, proximity-based communications in future wireless networks. In this communication type, two closely located user equipments (UEs) are eligible to bypass the cellular base station (BS) and effectively communicate in a direct mode \cite{3gpp.36.843}. The integration of D2D in emerging 5G systems is expected to be beneficial for mobile stakeholders, mainly by unlocking new applications related to services based on proximity, vehicular-to-vehicular (V2V) communications and other commercial opportunities that D2D can offer \cite{Tehrani2014}. In addition to that, there are also radical benefits in the overall network operation: first, the UEs' proximity entails increased data rate performance, low energy consumption as well as low transmission latency. Also, the reuse ability of the cellular and D2D links to simultaneously utilize the same radio resources translates to noteworthy spectral efficiency \cite{Lei2012}. Thus, because of the limited cellular resources and the ongoing proliferation of simultaneous data requests that need to be satisfied, the underlay notion of D2D communication is preferred due to its resource efficiency attribute.

In parallel, the virtualization of wireless radio resources has arisen as a promising solution to encounter the ongoing increasing data demand in today's and emerging future networks. Wireless resource virtualization (WRV) is currently emerging as a disruptive technology that offers significant benefits to different networks and service providers (N\&SPs) \cite{Kalil2014} as well as enabling vertical industries to create their own wireless network. Briefly, other than the fact that co-existing networks, which will be called tenants in the sequel, are able to share the substrate physical infrastructure that entails reduced capital and operational expenditures (CAPEX and OPEX, respectively), WRV ameliorates the utilization of radio resources via sharing them among the different N\&SPs \cite{Liang2015}. Hence, the exploitation of the wireless virtualization and network programmability merits on top of the integration of D2D paradigm can lead to improved network performance in terms of spectrum efficiency as well as overall network performance. 

However, the advent of the data-driven era brings in a number of challenges mainly due to the resulting cell and user densification \cite{Cisco}. Among all, a prevalent problem that is expected to attract not only academic but also industrial interest is that of direct communication between users that are subscribed to different mobile network operators (MNOs). A solution to this problem can create a fertile ground for introducing new business models that will fully leverage the D2D potentials. Technically, the weight should be primarily put on defining how the involved MNOs will coordinate their spectrum to satisfy their subscribers' quality of service (QoS) requirements. In the case of single MNO, underlaying D2D links are allowed to utilize the licensed cellular spectrum that is provided by the operator. On the other hand, in the case where two devices belong to two different operators, it needs to be decided which resources from which MNO will be utilized to realize the D2D connection. Therefore, the principal aim is to support a significant number of direct connections along different network operators while at the same time respect the performance of cellular users as well as the overall welfare of the system.

 In the next subsection, we delve further into the above issues as with respect to resource sharing across different tenants of the substrate infrastructure.

\subsection{Network Function Virtualization (NFV) \& Service creation}
\label{subsec:NFV}
\begin{figure}
  \includegraphics[width=0.48\textwidth]{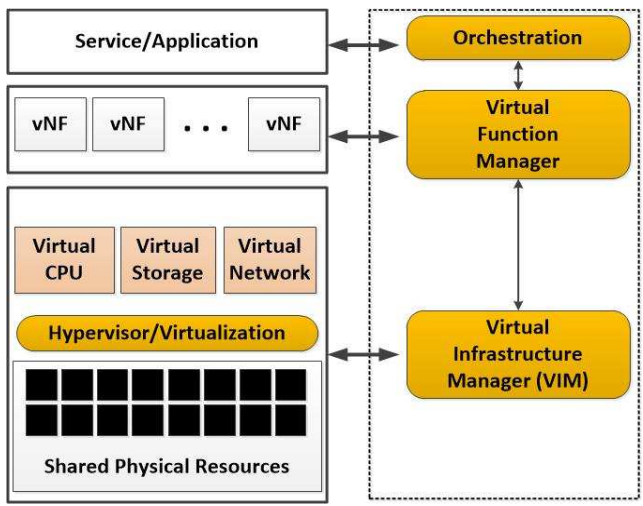}
\caption{Network Function Virtualization architecture following the ETSI framework.}
\label{fig:nfv}       
\end{figure}

In envisioned future virtualized and programmable 5G wireless network architectures, different tenants (virtual network providers) will be sharing the physical (substrate) network resources using a combination of Software Defined Networking\footnote{www.opennetworking.org} (SDN) and Network Function Virtualization\footnote{www.etsi.org/technologies-clusters/technologies/nfv} (NFV) architectures. The core idea behind NFV is to capitalize on virtualization technologies to  decouple physical network equipment from the services or functions that run on top of them \cite{NFV_overview}. Under the NFV framework, a network service  can  be  decomposed  into  a set of virtual network functions (vNFs) which are implemented in software and are able to run in general purposed hardware where they can be dispatched on demand. An overview of the NFV architecture is shown in figure \ref{fig:nfv} which conforms to the ETSI NFV framework. As depicted in this figure, a service request will be handled by the Orchestrator which will then inform the Virtual Function Manager about which vNFs are required to be activated for this specific service, whereas the actual physical resources for the vNFs will be handled by the Virtualization Infrastructure Manager (VIM). The above defined policies for the service creation will be distributed using the SDN controller (based for example on OpenFlow\footnote{Open Networking Foundation, OpenFlow Switch Specification Version 1.3.2 
April 25, 2013}). 

\begin{figure}
\centering
  \includegraphics[width=0.4\textwidth]{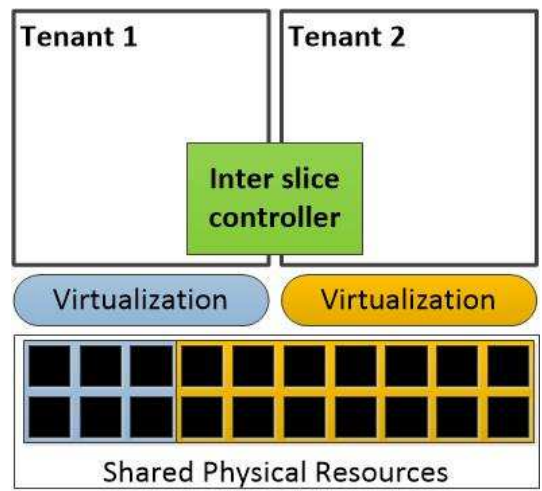}
\caption{A cross tenant communication entity that would allow efficient use of resources for users belonging to different tenants.}
\label{fig:cross_tenant}       
\end{figure}

Since each tenant will be allocated a slice of the available network resources (including also spectrum), mobile users that will require device-to-device communication from different tenants will be allocated resources (resource blocks) from the device which is originating the communication. This operation might lead to inefficient usage of the tenant's available resources in the long run. We therefore propose the use of an inter-slice coordinator that will allow for optimal usage of multiple tenants' resources in the case where the communication is taking place between users subscribed to different tenants. An illustrative example of an inter-slice controller is shown in figure \ref{fig:cross_tenant}, for the case of two tenants. Such cross-tenant orchestration would allow a more efficient use of the available physical resources per tenant. The concept of inter-slice coordination is being developed within the EU 5G-PPP 5G-NORMA project where the key motivation is to replace single RAN's networked entities by a network slice with a graph of programmable network functions\footnote{Mark Doll, 5G NORMA ``A Novel Radio Multi-service adaptive network Architecture for the 5G era", \emph{$1^{st}$ Sino-Europe 5G  Workshop}, November 2015, Beijing, China}. A cross-tenant controller should be a trusted entity since in order to optimize the overall performance tenants will have to provide intra-slice topological information to the controller which might include, inter alia, the number and location of users in each slice. Depending on the actual implementation, the cross-tenant controller can be considered as a broker that runs by the substrate network provider, which can be deemed as a trusted element. Note, there is an incentive for all tenants to cooperate since overall network performance is increased; however, trusted entities in virtualized architectures is a topic well beyond the scope area of this paper. Based on that fundamental assumption, the aim of this paper is to quantify the potential achievable gains enabled by such a cross-slice controller.

\subsection{Contribution and structure}
In this work, we assume that separate network slices allow for multi-tenant D2D discovery and session initiation procedures with similar techniques as the ones defined in \cite{patent}, where a protocol is designed to permit the inter-operator D2D communication. Without loss of generality, we consider two tenants and for each tenant we assume a number of subscribed cellular and D2D UEs (CUEs and DUEs, respectively), randomly distributed in a typical hexagonal cell layout. Each DUE is considered to be communicating in a maximum allowed distance with a peer that belongs to different tenant. Based on this topological modelling, we propose an integer linear programming (ILP) optimization framework that aims at maximizing the sum-rate performance of the involved inter-tenant D2D links while retaining the cellular UEs' QoS requirements of the involved tenants above a predefined performance threshold. To the best of our knowledge, this is the first paper that deals with the inter-tenant D2D communication optimization in virtualization-enabled networks. 

The structure of this paper is as follows: Section \ref{sec:literature} quotes a number of significant, closely related work on the issue of D2D virtualization. In Section \ref{sec:sys_model}, the system model that considers the integration of inter-operator (inter-tenant) D2D communications is initially described. Then, we proceed by formulating the sum-rate maximization problem for this communication paradigm. Section \ref{sec:results} illustrates the performance that is achieved based on this optimization methodology and compared to legacy techniques. Finally, concluding remarks are provided in Section \ref{sec:conclusion}.

\section{Closely related work}
\label{sec:literature}
In order to efficiently leverage the radical merits that D2D offers, the interference patterns that are being developed due to the resource reuse need to be limited. To this direction, several works within the literature have been elaborating on devising efficient resource allocation and mode selection techniques for D2D communications that mainly aim at improving network throughput as well as spectrum efficiency \cite{Asadi2014}. Optimal usage of the available radio resources  is a well-known nonlinear NP-hard problem that becomes even more complex with the integration of D2D communications that underlay the cellular network \cite{Zhang2013}. For this reason, a common practice has been to devise low-complexity, heuristic algorithmic solutions or to relax part of the constraints (e.g. power or resource block allocation related) and propose sub-optimal techniques for D2D-aware resource allocation \cite{Zhang2013}, \cite{Belleschi2011}.

So far, the problem of inter-operator D2D communication is inadequately explored, thus it needs to be carefully encountered in order to harvest the business dynamics of this specific communication type. To the best of our knowledge, the only existing work on this particular topic is \cite{Cho2015}. Therein, the authors propose the allocation of inter-operator D2D communications over dedicated licensed radio resources (overlay D2D) which the different operators have to negotiate between each other about the amount of spectrum that they will finally dedicate. They formulate this problem as a game between two distinct mobile network operators and decide about the offered spectrum with a best response method that runs in a sequential manner. Compared to it, our approach differs in that we consider a virtualized RAN infrastructure where inter-operator D2D links can utilize the whole available spectrum (underlay as opposed to overlay) to achieve  
 efficient resource sharing other than sum-rate maximization.
 
However, even the dynamics of intra-operator D2D communications in RAN virtualized ecosystems are barely explored in up-to-date literature. One of the first efforts towards this direction is the work in \cite{Cai2014}, where the authors address the problem of network state information (NSI) imperfectness in virtual wireless networks and resource allocation for the software-defined D2D connections. They devise a discrete stochastic optimization formulation to the problem of resource sharing given imperfect NSI and, then, proceed with the introduction of stochastic approximation algorithms for both static and varying channels resource manageability. Further, our work in \cite{Vlachos_Globecom} considered the virtualization of the resources offered by different mobile virtual network operators (MVNOs) in order to support and improve the performance of intra-MVNO D2D connections in the uplink scenario. The problem was formulated as an ILP sum-rate maximization problem, based on the constraint that the allocated resources per D2D should be contiguous. Heuristic proposed methods were also included as low-complexity solutions. 

Virtualization of the core as well as the radio access network is envisioned as the de-facto way forward for 5G networks since it can provide higher degree of flexibility to the mobile network operator, whilst with a careful design it can reduce overall network cost \cite{virtualization1}, \cite{virtualization2}. A preliminary study as with respect to use cases and requirements has also been defined within the 3GPP \cite{3gpp_requirments}. Also, important architectural aspects have also been discussed in order to support such advanced mechanisms \cite{3gpp_architecture}. Finally, under the assumption of a virtualized mobile network, the work in \cite{d2d_virtualization} considers the issue of resource allocation for D2D nodes via a non-linear optimization framework but does not consider the issue of inter-tenant D2D resource optimization.     

\section{System model}
\label{sec:sys_model}

\subsection{Preliminaries}
\begin{figure}
 \includegraphics[width=0.48\textwidth]{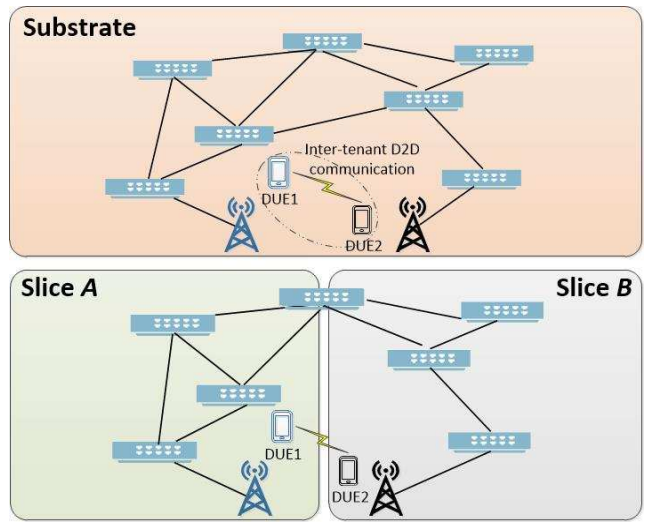}
\caption{Illustration of network slicing and inter-tenant D2D communication.}
\label{fig:topology}  
\end{figure}

The studied hexagonal cell area consists of a center-located RAN virtualized BS equipped with omni-directional antennas and a number of cellular users and D2D links uniformly distributed. Part of the distributed cellular and D2D users are assumed to belong to a specific tenant and are served by its designated slice, whereas the rest of them are subscribed to a second tenant, hence, a separate slice is dedicated to serve them. Note that, without loss of generality, we hereafter assume the existence of two tenants. Figure \ref{fig:topology} depicts the described scenario where intra and inter-slice/tenant D2D communications can take place. The establishment of  D2D communication is out of the scope of this paper. Briefly, intra-operator D2D session setup is carried out by the session initiation protocol (SIP) discussed in \cite{Doppler2009}, whereas the establishment and realization of the inter-operator D2D connection is detailed in \cite{patent}.

%
%

As already mentioned in subsection \ref{subsec:NFV}, each tenant is assigned with a slice that will provide, inter alia, spectrum allocation in order to fulfill the expected demand from the serving users. Quantitative, this translates to a number of resource blocks (RBs) which constitute the available resource pool of the users that are subscribed to specific tenant. Considering the legacy procedure, in order to support a D2D link between two users (intra or inter-operator), the resources used for it are allocated only from the RB pool that corresponds to the user that inaugurates the direct communication. However, in this work, we leverage the ability of an inter-slice manager to fuse the available RB pools of multiple tenants in order to enlarge the spectrum availability for inter-operator D2D links (figure \ref{fig:cross_tenant}).

Following the principles of D2D cellular spectrum reusability in the underlay notion, D2D users are able to utilize the resources of multiple cellular users simultaneously \cite{Wang2011}. However, it is important to ensure that the cellular transmissions whose resources are being reused by a D2D pair satisfy their QoS minimum requirements. For ease of comprehension, we presume that cellular users primarily occupy one but orthogonal radio resource based on the LTE specifications. 

\subsection{Problem definition}
\label{subsec:problem_def}
The reason why we consider this multi-tenant unified RB pool is to increase the resource efficiency for cross-tenant D2D links which are expected to be a significant part of future network connections. This will not only lead to effective usage of the available spectrum, but also improve the overall network performance by potentially increasing throughput and reducing interference. To this direction, an integer linear programming (ILP) optimization solution is proposed to maximize the sum-rate for inter-slice D2D links by respecting at the same time the cellular transmissions' performance not to degrade below a predefined threshold. 

Before we formulate the D2D sum-rate optimization problem, the following sets need to be defined:  
\begin{itemize}
\item $\mathcal{I}$ is the set of distributed cross-tenant D2D links; $\mathcal{I} = \{1, 2, ...., I\}$.
\item $\mathcal{C}$ is the set of cellular UEs; $\mathcal{C} = \{1, 2, ...., C\}$.
\item $\mathcal{N}$ is the set of tenants; $\mathcal{N} = \{1, 2, ...., N\}$.
\item $\mathcal{K}$ is the set of available resources; $\mathcal{K} = \{1, 2, ...., K\}$.
\end{itemize}
$\mathcal{C}$ contains all the cellular users that belong to different tenants and are consequently served by separate slices It can be represented as follows: $\mathcal{C} = \mathcal{C}_1 \cup \mathcal{C}_2 \hdots \mathcal{C}_N$, where $|\mathcal{C}_j| < C, ~\forall j \in \mathcal{C}$. 
Similarly, regarding the fused set of resource blocks (RBs) $\mathcal{K}$, it consists of all tenants' radio resources, so it can be written as $\mathcal{K} = \mathcal{K}_1 \cup \mathcal{K}_2 \hdots \mathcal{K}_N$, where $|\mathcal{K}_n| < K, ~\forall n \in \mathcal{N}$.

Further, we need to introduce the binary decision variable that indicates if a D2D link $i \in \mathcal{I}$ utilizes a specific RB $k$ that belongs to one of the tenants' available resource pool. This can be mathematically represented as: 
\begin{equation}
\label{eq:dec_var}
x_{ink} = 
\begin{cases}
1, & \text{if D2D link } i  \text{ uses RB } k \text{ of tenant } n \\ \\
0, & \text{otherwise} .
\end{cases}
\end{equation}

We further proceed with some important system model admissions to pave the way for the problem formulation. First, the path-loss is modeled as follows:
\begin{gather}
PL_\text{D2D} = 148 + 40\log_{10}d \\
PL_\text{CUE} = 128.1 + 37.6\log_{10}d
\end{gather}
for D2D pairs and cellular users, respectively \cite{3gpp.36.814}, \cite{Pang2013}. Parameter $d$ stands for the Euclidean distance and is expressed in kilometers (Km).
Additionally, the signal-to-interference-plus-noise-ratio (SINR) at the D2D receiver of link $i$ that uses RB $k = k_d$ of tenant $n$ needs to be satisfied. If we denote by $\gamma_{ink_d}$ this value which translates for the receiver's need to correctly decode transmitted packets,  this constraint can be practically expressed as follows:
\begin{equation}
\gamma_{ink_d} = \dfrac{ h_{ii}^{nk_d} P_d}{\sum_{n \in \mathcal{N}} \sum_{k \in \mathcal{K}} x_{ink}h_{ci}^{nk} P_c+I+\sigma^2} \geq \gamma_\text{th}
\end{equation}
where $h_{ii}^{nk_d}$ is the link gain (path-loss and slow fading dependent) of the $i^\text{th}$ D2D pair, and $P_d$ is the transmission power of the D2D transmitter over this RB. In the denominator, $h_{ci}^{nk}$ expresses the link gain between the transmitting cellular user equipment (CUE) $c$ and the receiver of D2D link $i$ when using RB $k$ of tenant $n$ (CUE-D2D interference will be developed when $k = k_d$ and $x_{ink} = 1$). As it will be mentioned in the sequel, we consider that D2D links will be using orthogonal RBs among each other, i.e. the received interference of the D2D links will be only deriving from cellular UEs (and vice versa). Lastly, $\sigma^2$ denotes the lump sum power of background/thermal noise and $I$ the co-channel interference from other cells (if existent). In that case, we assume that inter-cell interference can be controlled via the application of powerful inter-cell interference cancellation (ICIC) techniques, thus, we are focusing on a single-cell scenario ($I = 0$) where the main part of interference (i.e. intra-cell) is effectively captured. 

The SINR threshold ($\tilde{\gamma}_\text{th}$) needs to be also satisfied for the cellular transmissions that utilize the same RB (e.g. $k = k_c$) with a D2D link $i$ during the uplink session. This constraint can be written as follows:
\begin{equation}
\dfrac{h_{cb}^{nk_c} P_c}{\sum_{i \in \mathcal{I}} \sum_{n \in \mathcal{N}}\sum_{k \in \mathcal{K}} x_{ink}h_{ib}^{nk} P_d+I+\sigma^2} \geq \tilde{\gamma}_\text{th}
\end{equation}
where $P_c$ is the transmission power of a cellular user, $h_{cb}^{nk}$ is the link gain between the CUE $c$ that belongs to tenant $n$ and its associated BS $b$ when using RB $k$, whereas $h_{ib}^{nk}$ accounts for the link gain between the D2D transmitter of link $i$ and the BS $b$ that transmit/receive over the same channel $k$.

Considering the above definitions, the achievable rate for D2D link $i$ that utilizes RB $k$ of $n$ tenant can be calculated according to the well-known Shannon capacity formula:
\begin{equation}
\label{eq:shannon}
r_{ink} = B_\text{RB} \log_2 \left ( 1 + \gamma_{ink} \right )
\end{equation}
where $B_\text{RB}$ is the LTE-based resource block bandwidth (180 kHz) and $\gamma_{ink}$ is expressed in power ratio. 

Lastly, even though we focus on the uplink scenario where communications happen according to the SC-FDMA principles, we herein consider that the RBs allocated per user can be non-adjacent (\cite{Wang2015}) as the evolution of LTE towards 5G systems will eventually enable fully non-contiguous allocation. Considering this, we will practically provide an upper bound of the D2D-based rate performance.
Following the previous admissions, the sum-rate maximization problem for cross-tenant D2D communications can be formulated as follows:

\begin{subequations}\label{eq:optimization}
\begin{alignat}{2}
   & \underset{x}{\text{max}} \sum_{i \in \mathcal{I}} \sum_{n \in \mathcal{N}} \sum_{k \in \mathcal{K}}  r_{ink} x_{ink}  \tag{\ref{eq:optimization}} \\
\text{s.t.} &    ~ \sum_{i \in \mathcal{I}} \sum_{n \in \mathcal{N}}\sum_{k \in \mathcal{R}} x_{ink}h_{nk}^{ib}P_{d} \tilde{\gamma}_\text{th} \leq \nonumber \\ &- \Big( \tilde{\gamma}_\text{th}(W + I) - h_{nk_c}^{cb}P_{c} \Big),  \forall c \in \mathcal{C}, ~ k_c \in \mathcal{K} \label{eq:con1}\\
			 &     \sum_{n \in \mathcal{N}} \sum_{k \in \mathcal{K}} r_{ink} x_{ink} \geq r_i^{\text{th}}, ~ \forall i \in \mathcal{I} \label{eq:con2}\\
            &     \sum_{n \in \mathcal{N}} \sum_{k \in \mathcal{K}} x_{ink} \geq 1, ~ \forall i \in \mathcal{I} \label{eq:con3} \\
            &     \sum_{n \in \mathcal{N}} \sum_{k \in \mathcal{K}} x_{ink} \leq L_{\text{max}}, ~ \forall i \in \mathcal{I} \label{eq:con4} \\
            &     \sum_{i \in \mathcal{I}} x_{ink} \leq 1, ~ \forall n \in \mathcal{N}, \forall k \in \mathcal{K} \label{eq:con5} \\
            &     x_{ink} \in \{0,1\}, ~ \forall i \in \mathcal{I}, ~ \forall n \in \mathcal{N}, ~ \forall k \in \mathcal{K} \label{eq:con6}.
\end{alignat}
\end{subequations}
Constraint \eqref{eq:con1} ensures that each cellular transmission's SINR doesn't fall below a predefined value  $\tilde{\gamma}_\text{th}$, whereas \eqref{eq:con2} guarantees the minimum rate requirement for each D2D link $i \in \mathcal{I}$. Constraints \eqref{eq:con3} and \eqref{eq:con4} account for the radio resource allocation of each D2D link; the former ensures that each D2D pair will be assigned with at least one RB to satisfy its transmission needs, whereas the latter upper bounds the resources used by each link to $L_\text{max}$ to avoid any resource deficiency for some D2D UEs (DUEs). Then, the restriction that each RB can be used by only one D2D link is realized by \eqref{eq:con5}. Finally, \eqref{eq:con6} denotes the binary nature of the decision variable.

Finally, it is obvious that the overall rate achieved by a D2D link $i$ is $ r_i^\text{tot} = \sum_{n \in \mathcal{N}} \sum_{k \in \mathcal{K}}  r_{ink} x_{ink}$ and depends on the value assignment of the decision vector $\mathbf{x}$ that solves this optimization problem.

\section{Numerical investigations}
\label{sec:results}
In this section, a set of evaluation results is provided to shed light on the performance of the proposed inter-tenant D2D sum-rate optimization problem compared to legacy approaches and heuristic solutions. 

\subsection{Compared methodology}
In this subsection, a number of different D2D-based resource allocation techniques for inter-tenant communications are briefly described. These techniques constitute the compared methodology through which the results to follow are produced.

\begin{enumerate}
\item \textbf{Inter-tenant optimal}: The proposed method was detailed in subsection \ref{subsec:problem_def}. As previously explained, it yields optimal sum-rate performance for inter-tenant D2D users via a powerful ILP solution that virtually fuses the provided to the tenants RB pools and orchestrates the links' resource assignment.
\item \textbf{Inter-tenant heuristic}: Complementary to the previous technique, a heuristic algorithm is proposed to seek for a low-complexity, near optimal solution for D2D users that belong to different tenants. One of its chief characteristics is that it tries to achieve a fairly balanced, inter-slice resource allocation by sequentially running for D2D receivers that belong to different tenants. Its resource assignment rationale is based on allocating the resource blocks that provide the best channel conditions to each D2D link (in a sorted way) following the aforementioned sequential mode. 
Herein, it has to be noted that in order
for some RB to be assigned to a D2D link, first, it must be an unallocated one (among D2Ds) and second, not to degrade the performance of the cellular uplink transmission that utilizes the same RB.
Then, the algorithm iterates over and over, until one of the following conditions is violated: (i) all D2D users reach their upper RB usage limit (i.e. $L_\text{max}$ used RBs), (ii) the fused RB pool is fully utilized by the active D2D transmissions, or (iii) DUEs' SINR requirements over the remaining RBs are not satisfied.
The explained method is outlined in Algorithm \ref{alg:heuristic}.
\item \textbf{Intra-tenant optimal}: With this technique, problem \eqref{eq:optimization} is decoupled into two separate resource allocation problems for the two different tenants. This means that each tenant solves separately the sum-rate optimization problem for its subscribed D2D users that initiate direct peer communications, based on its corresponding dedicated slice resources. Due to the restricted RB availability for the different tenants, this method, even though it is able to provide optimal sum-rate performance from each tenant's side, it is expected to provide a sub-optimal solution in overall.  
\item \textbf{Intra-tenant heuristic}: Depending on the number of DUEs that are subscribed to a specific tenant and initiate a number of inter-slice connections, this tenant is the one to provide the corresponding direct communications with the suitable RB pool to satisfy their transmission needs. To this end, each one of the tenants allocates resources to the corresponding D2Ds in a greedy and sorted manner according to best-given channel conditions. This method is similar to the inter-tenant heuristic approach but again is decoupled as it needs to be solved by each different tenant for the subscribed users.    
\end{enumerate}

\begin{algorithm}
\DontPrintSemicolon
\SetAlgoLined
\nllabel{label}
\caption{\textsc{Inter-tenant heuristic algorithm}}
\KwData{CUEs'-DUEs' location coordinates, $L_\text{max}$.} 
\medskip
$\ast$ \textbf{Assumptions} : \\
\begin{itemize}
\item Each $c \in \mathcal{C}$ is assumed to be served by a specific tenant $n \in \mathcal{N}$.
\item Orthogonal, round-robin based resource allocation for  each $c \in \mathcal{C}_1 \cup \mathcal{C}_2$ is applied. 
\end{itemize}
\medskip

$\ast$ \textbf{Inter-tenant D2D resource allocation steps} : \\
Step 1 : \textit{D2D RB assignment} \\
\medskip
\Repeat{$\big<$RB pool is fully used \textbf{OR}  $L_\text{max}$ RBs are assigned for all D2Ds \textbf{OR} no more RBs can be assigned due to SINR requirements' violation$\big>$}{
$i = 1;$ ~ \small $\triangleright$ D2D link identifier. \normalsize\\
\While{$i \leq |\mathcal{I}|$}{
\eIf{($ i\mod2 \equiv 1$)}{  
$\bullet$ $slice\_id = A$; ~ \small $\triangleright$ slice/tenant identifier. \normalsize \\
$\bullet$ $i_A = i$; ~ \small $\triangleright$ D2D pair where the receiver is subscribed to tenant A. \normalsize \\
$\bullet$ Find the D2D-CUE combination that results in the maximum possible SINR for the D2D link; \\
$\bullet$ Allocate CUE's assigned RBs to link $i_A \in \mathcal{I}$;  \\
$\bullet$ Remove allocated RBs from available resource pool $\mathcal{K}$;\\
$\bullet$ $i = i + 1$;}{
$\bullet$ $slice\_id = B$; \\ 
$\bullet$ $i_B = i$;  ~ \small $\triangleright$ D2D pair where the receiver is subscribed to tenant B. \normalsize\\
$\bullet$ Repeat the same procedure and update RB pool $\mathcal{K}$; \\
$\bullet$ $i = i + 1$;}
}
}
\bigskip
\footnotesize{* Note: ($i\mod 2$) ensures the sequential RB allocation of D2D users for the example of two slices.} \\
\bigskip
\normalsize 
Step 2 : \textit{Sum-rate estimation} \\
\medskip
$\bullet$ $ r_\text{tot} = \sum_{i \in \mathcal{I}} r_i $; ~ \small $\triangleright$ $r_i$ is the achieved rate $\forall i \in \mathcal{I}$.  
\label{alg:heuristic}
\end{algorithm}

\subsection{Simulation setup}
The considered system was modeled in MATLAB, following the LTE-A milestones and corresponding network parameters and standards. 
All the produced results derived after averaging over 1000 Monte Carlo simulations which have been executed on a \textit{Intel(R) Core(TM) i7-6500 at 2.50 GHZ and 8 GB RAM} machine.

Regarding the topology, it consists of a hexagonal single cell with randomly distributed cellular users and D2D links. Each D2D link consists of two users that are assumed to belong to different tenants. Also, the number of cellular users being subscribed to different tenants is varying and considers the tenants' disparities in terms of number of subscriptions. Without loss of generality, two tenants, \textit{A} and \textit{B}, with separate slices are considered. All simulation parameters are listed in Table \ref{tab:sim_parameters}.  

\begin{table}
\centering
\captionsetup{justification=centering}
\caption{Simulation Parameters}
\label{tab:sim_parameters}       
\begin{tabular}{lll}
\hline\noalign{\smallskip}
\textbf{Parameter} & \textbf{Value}   \\
\noalign{\smallskip}\hline\noalign{\smallskip}
Cell layout & Hexagonal grid \\
Number of tenants ($\mathcal{N}$) & \(2\) \\  
CUEs-D2Ds distribution & Uniform\\
Macro cell radius  & \(400\) m\\ 
Maximum D2D link range  & \(100\) m\\
Number of CUEs (\(C\)) & 50 \\
Number of D2D links (\(I\)) & [10,40] \\
D2D Path-Loss  model & \(148 + 40\log_{10}d\)  \\ 
CUE-BS Path-Loss  model & \(128.1 + 37.6\log_{10}d\)  \\ 
Maximum CUEs' power  & \(20\) dBm \\
Maximum DUEs' power  & \(15\) dBm \\
Maximum number of RBs ($L_\text{max}$)  & \(4\)  \\
Shadowing standard deviation  & \(8\) dB \\
Noise power spectral density & \(-174\) dBm/Hz \\
System bandwidth (\(BW\)) & \(10\) MHz\\
\noalign{\smallskip}\hline
\end{tabular}
\end{table}

\subsection{Results}

\begin{figure}
\centering
\includegraphics[width=0.5\textwidth, trim = .5cm .0cm .5cm 0, clip = true]{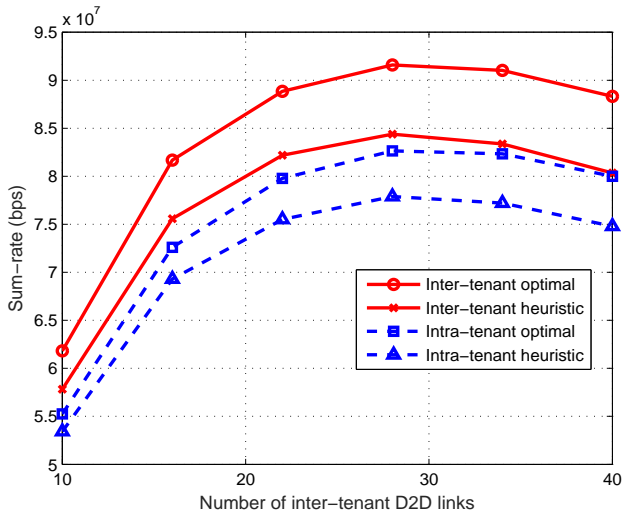}
\caption{Sum-rate comparison for inter-slice communications in relation to varying number of D2D links.}
\label{fig:1}
\end{figure}

Due to the load discrepancies and divergent number of subscriptions that might characterize the two or more tenants (either MVNOs or MNOs), the slice that each one is assigned with is expected to be different (i.e., having heterogeneous slices). Popular tenants can be normally allocated with more resources to serve the high number of subscribed users compared to less popular ones. To this end, we consider the case where a popular tenant (hereafter denoted as tenant \textit{A}) is allocated with double-sized RB pool to serve its subscribers. Initially, we assume that for both tenants all the radio resources are occupied by a number of cellular UEs according to a Round Robin scheduling. Figure \ref{fig:1} depicts the sum-rate performance for inter-tenant (inter-slice) D2D links in relation to varying number of them. On average, almost 11.3\% sum-rate gain is achieved by making use of the fused RB pools of the two tenants (inter-tenant optimal) compared to the case that a D2D link can be assigned resources only from the resource pool that belongs to the slice to which the user that initiates the direct communication is subscribed (intra-tenant optimal). The maximum performance gap among the illustrated scenarios is met in the case of 16 D2D links, where 12.5\% higher sum-rate is achieved with the inter-tenant optimization solution. Further, compared to the heuristic inter-tenant approach, the optimal solution is averagely 8.45\% better and gradually behaves better with the increase of inter-tenant D2D links' number. Also, the intra-tenant heuristic algorithm falls short compared to the above-mentioned approaches and it exhibits a maximum of more than 18\% sum-rate degradation in comparison to the optimal solution.  
Last, for all the considered approaches, the sum-rate drop that is observed in the two last cases (i.e., 30 and 40 D2D links) is explained by the interference increase to/from CUEs, as the resource availability gets more restricted.  

\begin{figure}
\centering
\includegraphics[width=0.5\textwidth, trim = .5cm .0cm .5cm 0, clip = true]{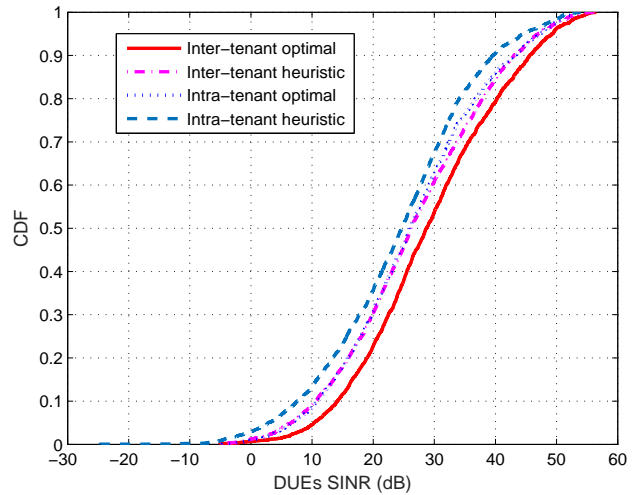}
\caption{SINR-based CDF for cross-tenant D2D links.}
\label{fig:1c}   
\end{figure}

Considering the same case study, the cumulative distribution function (CDF) of the achieved SINR values for the inter-tenant D2D links is represented in figure \ref{fig:1c}. Indicatively, in the $50^\text{th}$ percentile, the inter-tenant optimal solution's SINR for D2Ds is 29.1 dBs, whereas the corresponding values for the inter-tenant heuristic, the intra-tenant optimization and the intra-tenant heuristic are 26, 25.7 and 23.8 dBs, respectively. This can be interpreted as more than 2 times higher SINR power ratio compared to the inter-tenant heuristic approach. Last, the intra-tenant optimal and intra-tenant heuristic methods' estimated SINR is almost 2.2 and 3.4 times worse compared to the optimal value. 

\begin{figure}
\centering
\includegraphics[width=0.5\textwidth, trim = .5cm .0cm .5cm 0, clip = true]{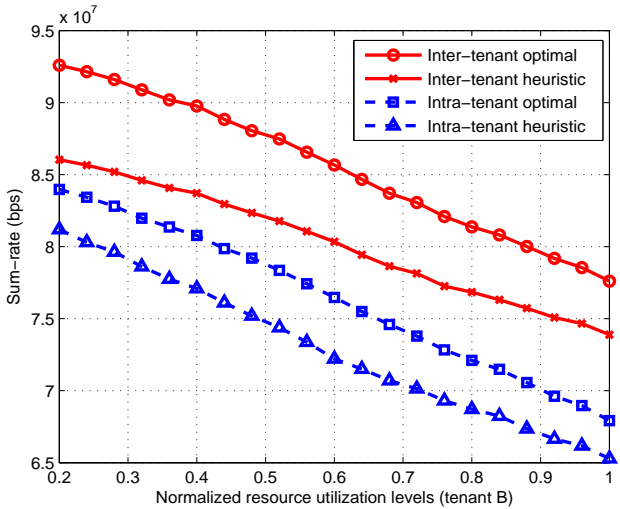}
\caption{Sum-rate performance in relation to RB utilization levels. The total number of D2D inter-tenant links is fixed to 20.}
\label{fig:2}  
\end{figure}

Then, we consider the scenario where the two tenants are characterized by the same RB availability but with different utilization levels (active cellular transmissions per case). To this direction, tenant's \textit{A} radio resources are supposed to be fully occupied by its subscribed CUEs, whereas tenant's \textit{B} resource block availability ranges from 20\% to 100\%. Figure \ref{fig:2} depicts the sum-rate performance of all compared methods in relation to the normalized resource utilization of tenant \textit{B}. As expected, while the resource occupancy increases, the sum-rate decreases for all methods as new interference patterns between cellular and D2D users arise. However, the performance gap between the inter-tenant optimal and heuristic technique lessens with the increase of the RB occupation levels for tenant \textit{B}, as opposed to the rest of the methods where the gap slightly increases. Quantifying the above observations, the inter-tenant proposed optimization formulation outperforms the rest of the algorithms in an average of almost 6.5\%, 12\% and 17.5\%, respectively. When both tenants' RB pools are fully utilized (reaching 100\% of resource utilization), all D2D links are reusing part of the cellular spectrum that CUEs occupy. In that case, the inter-tenant optimal solution achieves its peak sum-rate gain compared to the intra-tenant methodology; a 14.3\% improvement is observed over the intra-tenant optimal and 18.7\% over the related heuristic, respectively. This result can be deemed as highly interesting because the maximum gains take place when needed, i.e., during network congestion episodes.

\bgroup
\def\arraystretch{1.5}
\begin{table*}
\centering
\label{tab:runtimes}  
\caption{Algorithm running times}
\captionsetup{justification=centering}
\label{tab:run_times}       
\begin{tabular}{|l|l|l|l|l|}
\hline
\diagbox{\textbf{Algorithm}}{\textbf{Number of links}} & \textbf{I = 10}  & \textbf{I = 20}  & \textbf{I = 30}  & \textbf{I = 40} \\
\hline
\textbf{Inter-tenant optimal} & \(0.4516\) s & \(0.0465 \) s & \(0.0392\) s & \(0.0342\) s \\
\textbf{Inter-tenant heuristic} & \(0.0342\) s & \(0.0182\) s & \(0.0086\) s & \(0.0048\) s \\
\textbf{Intra-tenant optimal} & \(0.0786\) s & \(0.0412\) s & \(0.0351\) s & \(0.0332\) s \\
\textbf{Intra-tenant heuristic} & \(0.0282\) s & \(0.0204\) s & \(0.0169\) s & \(0.0055\) s \\
\hline
\end{tabular}
\end{table*}
\egroup

Further, the sum-rate performance of D2D users in relation to maximum link length is evaluated. By increasing the maximum allowable D2D link length, the sum-rate performance of all compared methods follows a decreasing trend. This is expected, as the increase of D2D link length implies higher SINR degradation not only due to path-loss and shadowing effects but also due to different emerging interference patterns. The performance gains become more clear for the largest values of link lengths. Therein, the inter-tenant optimization problem is almost 8.15\%, 18.9\% and 27.4\% better in terms of sum-rate when compared to inter-tenant heuristic, intra-tenant optimal and heuristic solutions, respectively. Also, on average, the inter-tenant heuristic method is the one that provides again the closest among all performance as it falls short almost 6\% in terms of throughput compared to the optimal one. Considering the rest, the optimal solution provides an average gain of almost 16\% and more than 22\% compared to the intra-tenant optimal and heuristic techniques.  

\begin{figure}
\includegraphics[width=0.5\textwidth, trim = .5cm .0cm .5cm 0, clip = true]{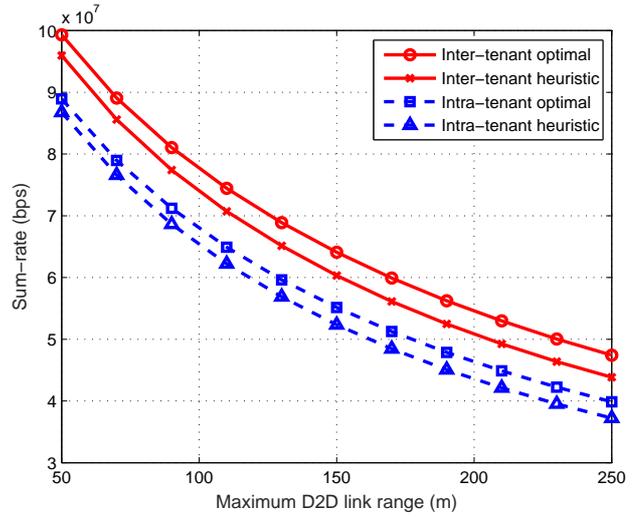}
\caption{Sum-rate performance in relation to the maximum allowable D2D link range.}
\label{fig:3}    
\end{figure}

Finally, in order to give a glimpse of their computational complexity, the running times of the aforementioned algorithms are listed in Table 2. It is shown that with the increase of the number of D2D inter-tenant links, the running times of all compared algorithms decrease. This can be explained by the fact that when less D2D links exist in the topology, the probability that all or many of them will utilize the maximum assignable number of resources ($L_\text{max}$) - in order to increase as much as possible their rate performance - raises. Consequently, the number of combinations for the orthogonal assignment of D2D links increases. Although the inter-tenant optimal solution is proven to be the most complex, its running time remains in acceptable levels.

\section{Conclusions}
D2D communications are expected to play a key role in emerging 5G wireless networks by unlocking a plethora of new proximity based applications while preserving the scarce wireless resources for the cellular users. Despite the fact that D2D communications received significant  attention there has been limited attention to the problem where the two UEs of a D2D communication belong to different service and/or network providers. Under the assumption of full virtualized core/access networks, the paper provides an optimization framework for efficiently use of virtualized resources across different tenants enabled by a cross-tenant controller. Via a wide set of numerical investigations it has been shown that significant throughput gains of over 10\%  compared to legacy solutions can be achieved for inter-tenant D2D communications. These results also reinforce the need for implementing a cross-slice coordinator, which can be considered as an extension to SDN/NFV frameworks, in order to efficiently utilize the scarce wireless resources across separate network slices.   
\label{sec:conclusion}

\section*{Acknowledgements}
This work, as a part of CROSSFIRE project, has received funding from European Union’s Seventh Programme for research, technological development and demonstration under grant agreement No. 317126. 
This work has also been performed in the framework of H2020-ICT-2014-2  project  5G  NORMA. The  authors  would  like  to acknowledge the contributions of their colleagues, although the
views expressed are those of the authors and do not necessarily
represent the project.


\end{document}